\documentclass[preprint,10pt]{aastex}

\newcommand{\grb}{gamma-ray burst}
\newcommand{\fl}{{f_{\rm lim}}}

\newcommand{\tm}{{t_{\rm mon}}}
\newcommand{\tb}{{t_{\rm break}}}

\newcommand{\thj}{{\theta_{\rm jet}}}
\newcommand{\thm}{{\theta_{\rm max}}}
\newcommand{\tho}{{\theta_{\rm opt}}}
\newcommand{\tmax}{{t_{\rm max}}}
\newcommand{\Omj}{\Omega_{\rm jet}}
\newcommand{\Omo}{\Omega_{\rm opt}}

\def\Omegajet{\Omega_{\rm jet}}
\def\Omegaopt{\Omega_{\rm opt}}
\def\rategrb{\dot N_{\rm grb}}
\def\rateopt{\dot N_{\rm opt}}
\def\tpeak{t_{\rm peak}}

\def\degsq{${\rm deg}^2$}
\def\tbreak{t_{\rm break}}
\def\fbreak{f_{\rm break}}

\begin{document}

\title{The difficulty in using orphan afterglows to measure
gamma-ray burst beaming}

\author{Neal Dalal, Kim Griest, and Jason Pruet}
\affil{Physics Department, University of California, San Diego, CA 92093}

\begin{abstract} 
If gamma-ray burst (GRB) emission is strongly 
collimated then GRBs occur
throughout the Universe at a rate much higher than is detected.  
Since the emission from the optical afterglow is thought
to be more isotropic than the gamma-ray emission,
it has been hypothesized that a search for orphan
afterglows (those without the triggering GRB) would allow strong constraints
to be placed on the degree of GRB collimation.
We show here that, within the context of leading models of
GRB jet evolution, measurement of the GRB beaming angle
$\thj$ using optical orphan searches is extremely difficult, perhaps
impossible in practice.  This is because in the leading model of GRB jets,
the effective afterglow beaming angle scales with the jet angle,
$\Omo\propto\Omj$ for small angles, and so the ratio of detected
orphan afterglows to 
GRBs is independent of the jet opening angle.  Thus, the number
of expected afterglow detections is the same for moderate jet angles 
($\sim 20^\circ$) as for arbitrarily small jet angles ($\ll 0.1^\circ$).
For nearly isotropic GRB geometry, or for radio
afterglow searches in which the jet has become non-relativistic, the
ratio of afterglows to GRBs may give information on collimation.
However, using a simple model we estimate the expected number of orphan
detections in current supernova surveys, and find this number to be
less than one, for all jet opening angles.
Even for future supernova surveys, the small
detection rate and lack of dependence on collimation angle appear to ruin
the prospects of determining GRB beaming by this method.  Radio
searches may provide the best hope to find the missing orphans.
\end{abstract}

\section{Introduction}
Gamma-ray bursts (GRBs) are observed at a rate of about one per day 
and are in some cases accompanied by optical afterglows that are useful 
in determining properties
of the burst such as its redshift.  Current models that attempt to
explain the enormous apparent energy release, the multiband spectra
and the temporal behavior of afterglow lightcurves typically
invoke a highly collimated relativistic jet which is beamed towards
the Earth.  The lightcurve from a highly collimated jet
is expected initially to decay as a low power of $t$ 
(e.g. flux $\sim t^{-1.1}$)
and then break to a steeper slope (e.g. flux $ \sim t^{-2.4}$) at late times
\citep{piran}.  A number of GRB afterglows
show evidence for just this behavior \citep{jets}.  
This has led \citet{frail01} to
conclude that the GRB emission is collimated into opening
angles between 3$^\circ$ and 25$^\circ$ with $\sim 4^\circ$ as an average.

One important feature of the jet model of GRBs is that
the vast majority of GRBs are beamed in directions away from the Earth and
therefore not observed in gamma-rays.  However, the afterglow
emission, occurring at much later time and at longer wavelength
should be beamed into a larger fraction of the sky
due to the decay of the jet Lorentz factor with time.  There should
therefore be afterglows that are observable when the associated
GRB is not \citep{rhoads97,loebradio,mrw,rhoads00}.  
As first suggested by \citet{rhoads97}, the ratio of the 
``orphan" afterglow rate to the GRB rate might allow measurement of
the GRB collimation angle $\thj$, via the equation 
$\Omegaopt/\Omegajet = \rategrb/\rateopt$,
where $\Omegajet$ is the solid angle subtended by the jet,
$\Omegaopt$ is the angle into which the afterglow is beamed, and $\rateopt$
is the efficiency corrected event rate for optical afterglows.
Thus, by counting the rate at which
orphan afterglows are detected, we could measure the average GRB jet
angle, and independently check other methods \citep{frail01} of
measuring GRB collimation.

Rhoads' method has already been applied several times to 
extant searches for type Ia supernovae and other optical transients.
Rhoads 
argued that the lack of afterglow detections by the Supernova
Cosmology Project and the High-$z$ Supernova Search roughly limited
$\Omega_{\rm opt}/\Omega_{\rm jet}\leq 100$, based upon a lack of
detections over 2 years' exposure.  
The continued non-detection of orphans since that time has been used
by several groups to set stronger collimation limits.
For example, \citet{rees} found $\Omegaopt/\Omegajet \leq 20$,
and \citet{rhoads01} quotes a transient search by Shaefer et al.
to claim $\Omegaopt/\Omegajet \ll 100$.
If optical afterglow emission is only mildly beamed (e.g. 
$\Omega_{\rm opt}\approx 1$), then this is
marginally inconsistent with the collimation factors ($\sim 500$)
claimed by \citet{frail01}, and could render untenable most solar mass
progenitor models, given the inefficiency of converting rest mass into
gamma-ray emission \citep{kumar}.

In this paper, however, we argue that current supernova searches place
no constraints upon GRB collimation.  One reason for this is that 
present surveys are too shallow,
narrow, and infrequent to detect significant off-axis emission.  Using
a simplified model of GRB afterglow emission and survey detection
efficiency, we show that present supernova surveys are generically
expected to detect (far) less than one orphan afterglow.
A deeper reason for the impotence of present surveys is
the fact that, within the context of the leading GRB afterglow
model \citep{jets,rhoads99}, the effective afterglow beaming angle
scales with the GRB jet angle, $\tho\propto\thj$.  
Because of this scaling, the ratio $\Omo/\Omj$ is unchanged even as
the GRB jet angle is varied from moderate angles  
($\thj\sim 20^\circ$) to arbitrarily small angles 
($\thj\ll 0.1^\circ$), and so the ratio of detected orphan afterglows
to GRBs cannot be used to measure $\thj$.  

We derive this
scaling below using a simplified model of GRB jet emission, however we
can understand this result in more general terms.
From relativistic kinematics, we know that the jet Lorentz factor
$\gamma$, beaming angle $\theta$, and observer time $t$ are related by
simple power laws, and as mentioned above the observed flux $f$ is a
simple power law in time.  Thus, the beaming angle is a (scale free)
power law in $\gamma$, $t$ and $f$.  In the limit $\gamma\gg 1$ the
only angular scale in the problem becomes $\thj$, so we expect the
effective beaming angle to scale as the jet angle, $\tho\propto\thj$.
This should break down in the nonrelativistic limit $\gamma\sim 1$.

This scaling of $\tho$ with $\thj$ severely hampers the ability of
optical orphan searches to measure or constrain GRB collimation.  In
fact, we argue in this paper that measurement of the GRB jet angle
may be impossible using either present or future optical supernova surveys.
The plan of the paper is as follows. In \S~2 we describe our simple
model of emission from a GRB afterglow.  In \S~3 we estimate
the afterglow detection rate using this simple model.
In \S~4 we discuss the results
of our calculations, and in \S~5 we give our conclusions, along
with some discussion and caveats.

\section{A simple model of optical afterglow emission}

GRB-alerted afterglows (``on-axis afterglows'') are now well
studied. In most cases, their time dependence can be characterized as a broken
power-law, $f\sim t^{-\alpha}$, with a power-law break
from $\alpha\approx 1.1$ to $\alpha\approx 2.4$
occurring at late times when the relativistic beaming angle
$\gamma^{-1}$ exceeds the jet opening angle $\thj$.  
Here $\gamma$ is the bulk Lorentz factor of the relativistic jet.
This break occurs at a time \citep{jets}
\begin{equation}
t_{\rm break}\approx 6.2(E_{52}/n_1)^{1/3}(\thj/0.1)^{8/3}\textrm{hr},
\end{equation}
where $E_{52}$ is the inferred energy of the ejecta assuming isotropic
expansion, in units of $10^{52}$ ergs, and $n_1$ is the density of the
surrounding ISM in cm$^{-3}$.  GRB optical emission has a wide range
in apparent brightness, ranging from 9th magnitude for the prompt
optical emission associated with GRB 990123, down to the limiting
magnitude $R\sim 23-24$ of follow-up searches for afterglows.
\citet{fynbo} have argued that most ($\sim 70\%$) afterglows are even
fainter than this, and have therefore eluded detection.

Observers of orphan afterglows, on the other hand, would see off-axis
lightcurves, which differ from on-axis lightcurves. 
The behavior of off-axis emission is governed by
relativistic effects, as well as the lateral
spreading and internal structure of the jet.  We will first consider 
the effects of relativistic kinematics, 
and then consider lateral spreading and jet
internal structure.  Since the internal structure and time evolution
of GRB jets is quite uncertain we will consider throughout two limiting
cases which we hope will bracket reality.

\begin{figure}
\plotone{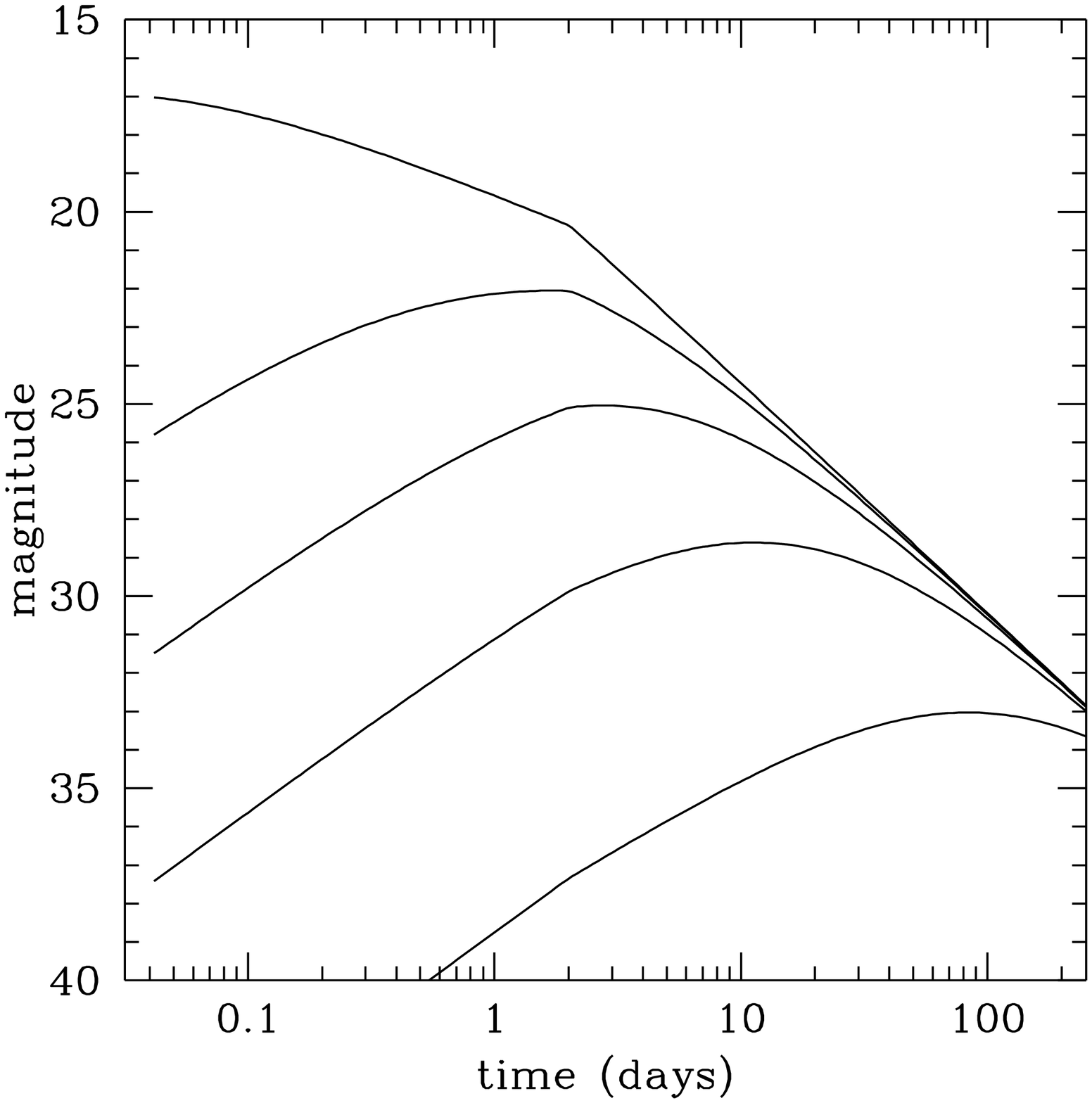}
\caption{Apparent magnitude of
a GRB afterglow as a function of time for various viewing angles 
as given by equation \ref{eqn:real}.
From top to bottom the curves are for viewing angles 
of $\theta = $0.01, 0.05, 0.1, 0.2, and 0.5 radians, respectively. 
The GRB is taken to be at a redshift of 1, have absolute magnitude
$M = -25$ at the break time of 1 day in the GRB rest frame
(corresponding to observed break time of $\tbreak=2$ days).
The jet opening angle is $\thj = 0.07$ radians, and it is assumed
that the flux is concentrated at the jet center and lateral spreading
of the jet is not included. 
Note that the
flux peaks near the time when $\gamma\approx\theta^{-1}$.  Before
the peak, the flux rises as roughly $t^{2m\mu-\alpha}$, 
and after the peak it decays as $t^{-\alpha}$.
\label{reallightcurve}}
\end{figure}

As a simple model, consider a radiating plasma moving relativistically
along the jet axis, with bulk Lorentz factor $\gamma$ relative to us,
and radiating isotropically in its own frame.  The measured power, as
a function of angle $\theta$ with respect to the jet axis in our
frame, is \citep{radbook}
\begin{equation}
\frac{dP}{d\Omega}=\frac{F}{\gamma^4(1-\beta\cos\theta)^4}
\end{equation}
where $F$ is the (isotropic) power per unit solid angle in the plasma
frame, and $\beta=(1-\gamma^{-2})^{1/2}$ as usual.  Let us assume that 
the Lorentz factor $\gamma$ and rest frame flux $F$ evolve with time
to give an observed on-axis ($\theta = 0$) lightcurve $f(t)$
matching the lightcurves observed for optical GRB afterglows, i.e. 
$f(t)\sim t^{-\alpha}$.  
For early times before the break $\alpha = 1 - 1.2 \approx 1.1$
while at late times after the break $\alpha \approx 2.4$.
Since $\gamma \propto t^{-\mu}$, 
with $\mu=3/8$ before the break and $\mu = 1/2$ after the break
\citep{jets}, this
means $f\propto\gamma^{\alpha/\mu}$.  This then requires 
$F\propto\gamma^{\alpha/\mu} (dP/d\Omega(\theta=0))^{-1}$.
With our assumption of rest frame isotropy, we obtain
the flux seen by off-axis observers as a
function of angle $\theta$ and Lorentz factor $\gamma$,
\begin{equation}
f(t)=\fbreak\left(\frac{t}{\tb}\right)^{-\alpha}
\left(\frac{1-\beta}{1-\beta\cos\theta}\right)^m.
\label{eqn:real}
\end{equation}
Here $\fbreak$ is the flux seen by an on-axis observer at the break time
$\tbreak$, and we have generalized the exponent 4 to the parameter $m$
for later use.  
This function with $m=4$ is plotted as a function of $t$ for
several values of $\theta$ in Figure \ref{reallightcurve}.
Several features follow from equation \ref{eqn:real} and Figure
\ref{reallightcurve}.  First, 
the lightcurve seen at angle $\theta$ peaks at a time $\tpeak$ which
can be estimated by setting $\gamma\approx\theta^{-1}$.
Next, the flux rises to the peak roughly
as $t^{2m\mu-\alpha}$, and decays after peak as $t^{-\alpha}$.
Also, the flux is nearly independent of angle for
$\theta\ll\gamma^{-1}$; that is the various lightcurves all match
together at late times when the viewing angle becomes small 
relative to the beaming angle. Finally, when 
$\gamma< \theta^{-1}$, an off-axis viewer at $\theta>\thj$ should see
comparable flux to an on-axis observer inside $\thj$, and the off-axis
afterglow lightcurve should decay with roughly the same slope as the
on-axis afterglow.  

Note that off-axis ($\theta>\thj$) afterglows do not peak until
after the on-axis power-law break, which typically occurs 
hours to days after the GRB \citep{frail01}.
This implies that orphan afterglows, even at peak
brightness, should be far fainter than their GRB-triggered siblings,
and thus should be much more difficult to detect.
We can roughly estimate how much fainter as follows.  After the break,
the afterglow flux decays as $t^{-2.4}$, while the Lorentz factor
decays as $\gamma\propto t^{-1/2}$ \citep{jets}, so the flux scales as
$\gamma^{4.8}$.  For a reasonable estimate of the jet angle, we would
like to detect orphans at large viewing angles, say $\theta\approx
0.2-0.3$.  This means detecting emission at times when the Lorentz
factor has decayed to $\gamma\approx 3-4$.  Taking 
$\gamma\approx (4^\circ)^{-1}\approx 15$
at the break, we see that the flux is 7 or 8 magnitudes fainter by the
time $\gamma\approx 3-4$ than at the break.

This immediately shows why current supernova searches are not expected 
to detect orphan afterglows.  Typical limiting magnitudes for such
searches are near $R\approx 23$, which \citet{fynbo} claim is too
shallow to detect most on-axis afterglows, let alone far fainter off-axis
orphan afterglows.  This implies that
detectable orphan afterglows are mostly contained in a
cone of opening angle not much larger than $\thj$, so we do not
expect many more orphan afterglows to appear in SN surveys than
on-axis afterglows.  Additionally, we show in the next section 
that the relatively
infrequent monitoring characteristic of current SN surveys
(e.g. $\tm=21$ days) causes most bright orphans to be missed.

We shall quantify these broad statements in the next section with a
calculation of the detection rates expected in surveys, using this
simple model of off-axis afterglow emission.  Before proceeding, we
pause to note that we have, so far, neglected the effects of finite
jet width and jet spreading.  Both of these should
lead to greater off-axis emission than indicated by
equation \ref{eqn:real}.
To obtain an upper limit to the effects of the uncertain jet evolution
(and thus bracket the range of possible fluxes seen by off-axis observers), we
consider the following simple picture.
The effect of finite jet width on the off-axis lightcurve is
to point some of the emitting material more closely to the
observer's line of sight, and to point other emitting material further from
the observer's line of sight. A proper calculation of this effect would
involve specifying the emission profile in the jet and integrating
over the jet face \citep{woodsandloeb1999}, but in the spirit of
setting upper bounds on the detection rate, we can get a sense of
the effect by simpler means.  Consider the emission from the part of
the jet face most closely pointing to a fiducial off-axis observer at
$\theta$.  This material is moving at an angle $\theta-\thj$ from the
observer's line of sight, so we can approximate the flux from this
part of the jet face by replacing $\theta$ with $\theta-\thj$ in equation
\ref{eqn:real}.  We immediately see that the asymptotic scalings
remain unchanged (i.e. $t^{4-\alpha}$ and $t^{-\alpha}$), but that the
peak time changes, to $\gamma^{-1}\approx\theta-\thj$.  This makes
sense, because we roughly expect the flux to beam into an angle
$\sim\thj+\gamma^{-1}$.  This of course overestimates the total flux
and the true visible angle, because emission from the rest of the jet
face is less blueshifted and therefore less bright.  Also note that
the emission from other regions of the jet face should peak at later
times, so that we expect a broader lightcurve peak for off axis
viewers than on axis viewers.  This broadening of the peak is seen in
numerical simulations of relativistic jet hydrodynamics and
emission \citep{hydrojet}.

An upper limit on 
the effects of lateral spreading of the jet on the off-axis lightcurve
may be obtained by noting that if in the jet comoving rest frame the
emitting material is expanding laterally at speed $v$, then the apparent
angular size of the jet face becomes
$\thj=\theta_{{\rm jet},0}+(v/c)\gamma^{-1}$.
Then our previous discussion applies, merely replacing the constant
jet angle with this spreading jet size.
To bracket possible jet morphologies and 
evolution, for a lower bound on off-axis emission we use 
equation \ref{eqn:real}, which has all emission coming
from the jet center and no lateral spreading, and for an upper bound
on off-axis emission we use equation \ref{eqn:real}
with $\theta$ replaced by $\theta - \thj - \gamma ^{-1}$, which
has all emission coming from the nearest jet edge and lateral jet
spreading at the speed of light.

\section{Event detection rate}

Observing programs that aim to measure GRB collimation using orphans
will measure the rate of orphan afterglow detection and compare 
this to the
rate of GRB-triggered (on-axis) afterglow occurrence.  Optical
searches for orphan afterglows naturally piggyback on top of Type Ia
supernova searches, which repeatedly monitor fixed regions of the sky
and use differential photometry to detect transient optical sources.
Ideally one would like to detect every afterglow occurring in the
observed region, but some afterglows will be missed either because
they are too faint or because the monitoring is not frequent enough to
catch them.  Note, however, that if one can accurately estimate the
efficiency at which afterglows are found, then orphan searches may be
useful even if many afterglows are missed.

In this section, we
estimate the number of afterglows that an observing program
would detect and the properties (such as viewing angle from the center
of the jet) of those detected afterglows.  We can estimate these quantities
using the simple model of off-axis emission described in the previous
section.  First, we give a simple analytic calculation of the
detection rate, appropriate in the limit of frequent or continuous
monitoring, which captures the effects of jet behavior.  
To determine the effects of infrequent monitoring, we
then perform a Monte Carlo simulation of a simplified observing program.

\subsection{Analytic estimate}

Consider a perfect orphan afterglow search capable of
detecting any orphan afterglow above limiting flux $\fl$.
We would like to determine which viewing angles $\theta$ see apparent
fluxes above threshold, $f>\fl$, as a function of time $t$.
Using equation \ref{eqn:real} and
$\gamma = \thj^{-1} (t/\tb)^{-\mu}$, and taking $\gamma\gtrsim$few, we
see that $f>\fl$ for viewing angles $\theta<\thm$, 
where $\thm$ is given by
\begin{equation}
1-\cos\thm\approx \frac{1}{2}\thm^2\label{eqn:genl}
=\frac{1}{2}\thj^2\left(\frac{t}{\tb}\right)^{2\mu}\left[\left(
\frac{\fbreak}{\fl}\right)^{1/m}\left(\frac{t}{\tb}\right)^{-\alpha/m}
-1\right]
\end{equation}
Recall that before the break, $\mu=3/8$ and $\alpha\approx 1.1$, while
after $\tb$ we have $\mu=1/2$ and $\alpha\approx 2.4$.
To compute the effective afterglow beaming angle, we maximize equation
\ref{eqn:genl} with respect to time.  Let us call $\tmax$ the time at which
the visible angle is maximized.
Depending upon whether or not $\tmax$
coincides with the break time, the $\thm$ expression becomes
\begin{equation}
\theta_{{\rm max},0}=\thj\times\left\{
\begin{array}{cl}
\left[(\fbreak/\fl)^{1/m}-1\right]^{1/2}&\quad\tmax=\tb\\
\left[\frac{\fbreak}{\fl}\left(1-\frac{\alpha}{2\mu m}\right)^m\right]
^{\mu/\alpha}\left(\frac{\alpha}{2\mu m-\alpha}\right)^{1/2}
&\quad\tmax\ne\tb
\end{array}\right.
\label{eqn:nojetnospread}
\end{equation}

This discussion has neglected both the finite width of the jet and
the possibility of lateral spreading of the jet.  Including these effects
as discussed in \S~2 gives
\begin{equation}
\thm(z)=\thj+\left(1+\frac{v}{c}\right)\theta_{{\rm max},0}
\label{eqn:withjetandspreading}
\end{equation}
where $\theta_{{\rm max},0}$ is defined as in equation \ref{eqn:nojetnospread}.
We emphasize that this is an upper limit to the visible angle, and that a
more careful treatment would give a $\thm$  somewhere in
between that given by equation \ref{eqn:nojetnospread} and
equation \ref{eqn:withjetandspreading}.  Our previous
expression for $\theta_{{\rm max},0}$ should underestimate off-axis
emission, and so these two cases should bracket all possible jet emission.

In order to proceed, we next need to define the GRB properties.
There are large uncertainties here, and we expect these to introduce
substantial uncertainties 
in our estimates of the event rates.  
However, we argue that this will not affect our main conclusions.

First we follow \citet{wijers} and \citet{loebsn} 
in hypothesizing that the comoving
rate at which GRBs occur is proportional to the star formation rate, 
$R_{\rm SFR}$. 
For simplicity we approximate the
star formation rate, as measured by \citet{steidel}, as
\begin{equation}
\log_{10} R_{\rm SFR}(z)=\left\{
\begin{array}{cl}
Cz&z<1\\
C&1<z<10,
\end{array}
\right.
\end{equation}
with an arbitrary cutoff for $z>10$.
For the $\Omega_M=1,\ H_0=50$ km/s/Mpc Einstein-de Sitter universe
that \citeauthor{steidel} use, $C\approx 1$ appears adequate, and correcting
this to the flat $\Omega_M=0.3,\ \Omega_\Lambda=0.7,\ H_0=70$ km/s/Mpc
cosmology that we adopt gives instead $C\approx 0.75$.  We absorb
the overall normalization of $R_{\rm SFR}$ into our total event rate, which
we normalize so that the overall rate of GRBs is 
$\sim 666/$yr over the full sky \citep{batse}.
\footnote{
Note the number 666/yr includes both long and short GRBs, while
afterglows have been detected only for long bursts.  We conservatively
assume here, however, that all bursts give rise to afterglows; 
this will soon be testable with experiments such as HETE-II or Swift.}

Next we must make an assumption concerning the luminosity function
of GRB afterglows.  For simplicity we will assume that afterglows are
standard candles with absolute magnitude $M=-25$ at the time of
the lightcurve break.  This number is chosen so that $\sim 70\%$ of
well-followed GRBs do not show afterglows because the afterglows are
too faint at the time of observation, which we assume to be near the
break time. 
Of course, it is known that the dozen or so afterglows
with measured redshifts are widely distributed in absolute magnitude,
but $M=-25$ is a typical value.
Since the effective beaming angle depends strongly on the apparent luminosity
function (c.f. equation \ref{eqn:nojetnospread}), 
uncertainty here will lead to substantial uncertainty in the
event rate, as we discuss below.

One additional concern is the k-correction.
At high frequencies, the afterglow
spectrum is expected to behave as $F_\nu\sim\nu^{-\alpha/2}$
\citep{jets}, so as a function of redshift $z$,
$\fbreak(z)=\fbreak(0)\times[1+z]^{-\alpha/2}$.  

Finally, we assume that all GRBs have the same jet opening angle
$\thj$, the same power law indices $\alpha$ and $\mu$ specified
in \S~2, and the
same rest-frame break time $\tbreak \approx 1$ day. 
Note that here by rest-frame, we mean as viewed by an observer at the same
redshift, so that there is a $(1+z)$ correction.  We do not mean the
plasma rest frame time, for which there is an additional $\gamma^2$
correction relative to observer time.
Clearly there is a wide spread in the observed $\tb$'s and
power-law indices $\alpha$ for previous GRBs \citep{frail01}, and this
will lead to uncertainties of order a few in our final numerical results.

We can now compute the expected rate of orphan afterglow detections. 
Let us write the effective beaming (solid) angle as
$\Omega_b(z) = 4\pi(1-\cos\thm(z))$.  With our assumption of standard
candle afterglows, we need only specify the redshift distribution to
obtain the afterglow luminosity function.
Assuming as above that the \grb\ rate is proportional to the star formation
rate, we may write the
observed afterglow redshift distribution as
\begin{equation}
\frac{d{\dot N}}{dz}=A\frac{dV_{\rm co}}{dz}
\frac{R_{\rm SFR}(z)}{1+z}\Omega_b(z),
\label{eqn:dndz}
\end{equation}
where $dV_{\rm co}/dz$ is the comoving volume element, $z$ is the
GRB redshift, and $A$ is a normalization constant.
The k-correction is incorporated as above and we set $A$ as above by
assuming that all
on axis \grb s are bright enough to be detected; that is we set
$\thm=\thj$ in equation
\ref{eqn:dndz} and integrate over redshift giving
${\dot N}_{\rm GRB}=A\cdot\Delta\Omegajet
\int_0^\infty dz (dV_{\rm co}/dz) R_{\rm SFR}/(1+z) = 666$ yr$^{-1}$
over the full sky.

The calculation described above elucidates the overall behavior, 
but neglects effects which may
affect numerical predictions for the overall detection rate.
For example, we have extrapolated the scaling formulas for $\gamma$ to the
nonrelativistic regime where the scalings break down; fortunately
we will show that the regions of low $\gamma$ make a small contribution
to the integral for all but the deepest surveys. 
We have also assumed continuous monitoring, which is certainly
not the case in current SNe searches.  In order to take into account
the effect of imperfect monitoring
and to check the above analytic calculation
we ran a Monte Carlo simulation of a simplified observing program.

\subsection{Monte Carlo simulation}
We define an observing program by its limiting magnitude $m_{\rm lim}$
(and corresponding limiting flux $\fl$), by the time between exposures
of a given field $\tm$, and by the total survey exposure (number of
\degsq yr).  In an actual survey additional parameters and various experimental
efficiencies would also be taken into account but these are not crucial
for our considerations here.  
The observing program simulation is run for a given value of $\thj$
by distributing GRBs randomly throughout the Universe, proportionally
to the product of comoving volume and star formation rate, and at a random
viewing angle to the observer.  The lightcurve is started
at a random time and observed every $\tm$ days.  The afterglow is considered
to be detected if any of the observations find a flux above the limiting
flux of the observing program.  The total event rate and the distribution of
detected events as a function of redshift, observing angle, etc. is
thus easily found.  

\section{Results}

Now, all the elements are in place to calculate the detection rate and
average beaming factor for a given survey.  Let us first consider a Type
Ia supernova search \citep{hzss,scp}, with 
limiting magnitude of $R\approx 23$ and monitoring repeat
baseline $\tm\approx 21$ days.  Adopting an opening angle
of $\thj=4^\circ$ \citep{frail01} our  Monte Carlo calculation
gives an expected rate of $\sim$ 0.0008 deg$^{-2}$ yr$^{-1}$ 
if the flux is concentrated at the jet center and there is no lateral 
spreading, and $\sim$0.014 deg$^{-2}$ yr$^{-1}$ 
if lateral spreading is maximal and the flux is concentrated at
the jet edge.
%
So, taking an exposure of 10
deg$^2$ yr, the we predict between 0.008 and 0.14 detections
from previous Type Ia supernova surveys.  We note that in the limit of
very small $\tm$ the Monte Carlo results are close to the analytic
results, which finds 3 - 5 times more events depending on jet morphology.

Now, the important point to note is that these numbers, besides being
small, are fairly
independent of $\thj$.
The reason for this should be clear.  The only
dependence of equation \ref{eqn:dndz} on $\thj$ is the $\Omega_b(z)$
factor, which was the effective beaming angle as a function of
redshift.  Since $\thm\propto\thj$, $\Omega_b\propto\Omegajet$.
But since the total GRB rate scales like $\Omegajet^{-1}$ to
hold fixed the number of observed GRBs, this means that the expected
number of detected orphans is also held fixed.  So the non-detection
of orphan afterglows is not only consistent with opening angles of
$4^\circ$, it is consistent with almost {\it any} opening angle.

Next we consider a more powerful future observing program with
limiting magnitude $R_{\lim} \approx 27$, and frequent enough
monitoring that our analytic estimates are accurate.
Here the number of predicted afterglow detections can rise above
unity if many tens of square degrees are monitored.
The number of expected events is shown in Figure \ref{thetavariation} as
a function of the opening jet angle $\thj$.  
For opening angles ranging from moderate ($\sim 20^\circ$) to
arbitrarily small ($\ll 0.01^\circ$) there is essentially no change in
the predicted number of detections.  Note, however, that for large jet
angles ($\gtrsim 30^\circ$) there can be a significant decrement in the
predicted number of detected orphans, in part because the relativistic
scalings we have employed break down, and in part because the emission
begins to subtend the full sky.
However, it is clear that for the angles being considered in
the leading jet models, the ratio of GRB detection to afterglow detection
will give no information on the GRB collimation.

\section{Discussion and conclusions}

We have shown that within the framework of the leading
model for the late time GRB jet evolution, current optical
surveys cannot constrain GRB beaming. Part of the problem is that current
surveys are too infrequent, narrow and shallow.

\begin{figure}
\plotone{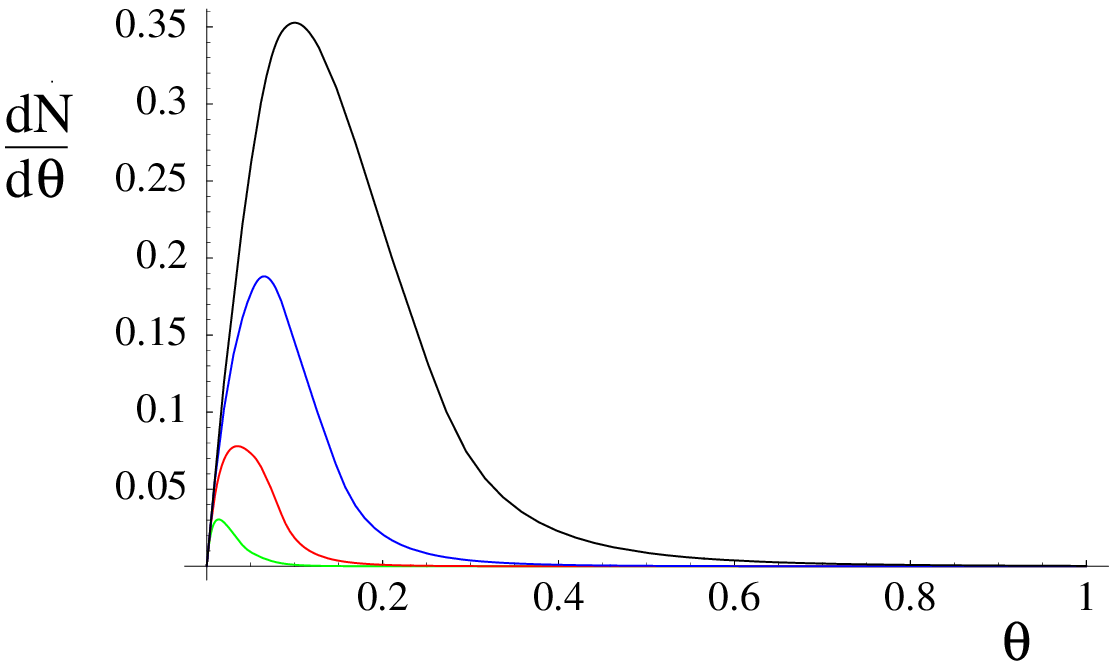}
\plotone{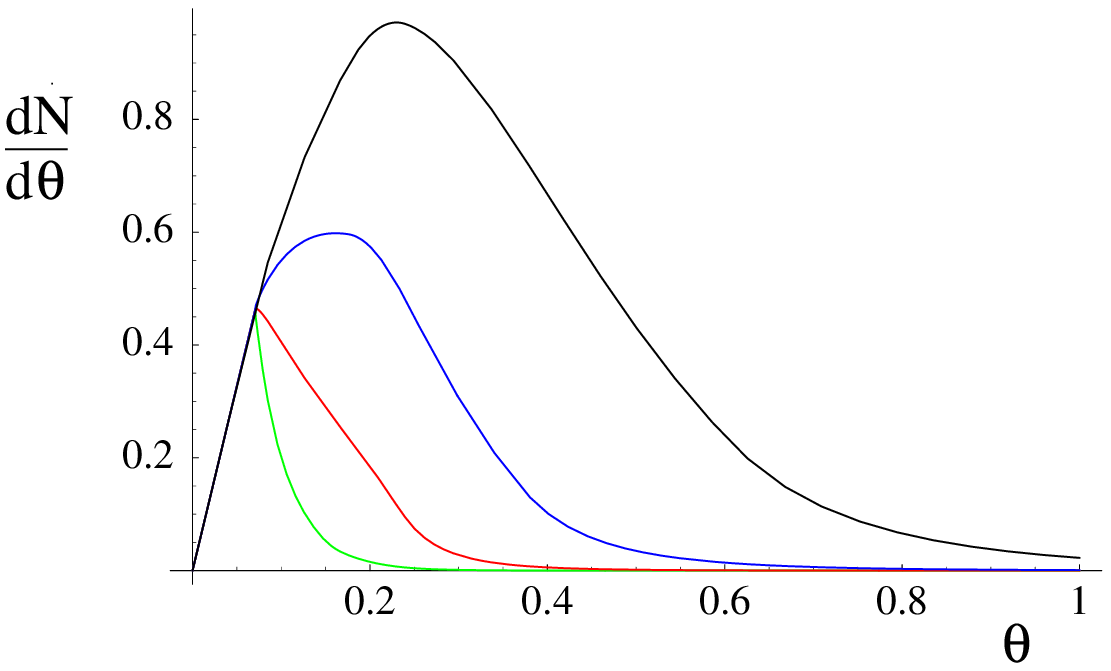}
\caption{Plot of $d{\dot N}/d\theta$, in units of events/yr/deg$^2$,
as a function of viewing angle $\theta$, for an assumed jet angle
$\theta_{\rm jet}=4^\circ$ and for our analytic approximation.
The different curves within each panel
show the effect of varying the limiting
magnitude.  From top to bottom, they are $m_{\rm lim} =$ 30, 27, 
24, and 21, respectively.
The different panels correspond to different assumptions of the
effects of jet width and lateral spreading.  In the first panel, off
axis emission is estimated using equation \ref{eqn:nojetnospread},
that is, it assumed to come only from the jet center.  In
the second panel, off-axis emission is estimated using equation
\ref{eqn:withjetandspreading} with $v=c$, corresponding to 
emission from the nearest jet edge and maximal spreading.
Note that $d{\dot N}/d\theta$ is independent of limiting magnitude for
$\theta<\thj$ when we include the finite jet width, since on-axis
afterglows in our model can be arbitrarily bright at early times, and
we assume continuous monitoring in our analytic approximation.
\label{dndtheta}}
\end{figure}

A more basic problem is that the effective afterglow beaming angle
scales as the jet angle for small ($<20^\circ$) jet angles, and so the
ratio of $\Omega_{\rm opt}/\Omega_{\rm jet}$ cannot be used to measure
$\thj$.  This conclusion depends on the simple scaling laws of flux
and Lorentz factor that apply during the relativistic phase of the
expansion. In order to see emission from large viewing angles, one
must wait until the jet has become only moderately relativistic and
therefore dim.
In Figure \ref{dndtheta} we plot the distribution of viewing angles
for several different limiting magnitudes. Only
the $R=30$ lightcurve has any sensitivity to large viewing angles, and
it is still dominated by small $\theta$.  
Finding optical afterglows this faint or fainter will be difficult due
to, for example, host galaxy domination.  

\begin{figure}
\plotone{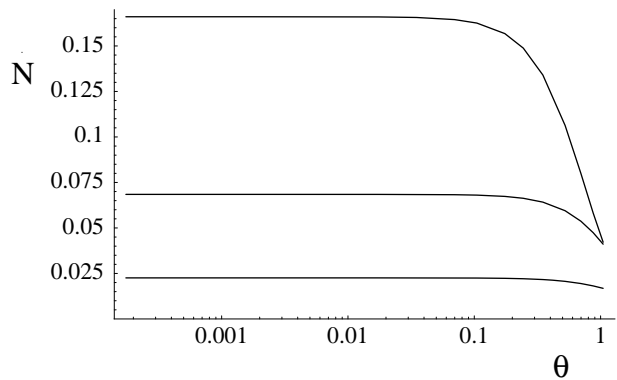}
\caption{Plot of total number of afterglows detected per year
per square degree as a function of jet angle $\theta_{\rm jet}$ in
radians for our simple approximation.  A limiting magnitude of
$R=27$ was assumed. The different curves show the effects of jet width
and lateral spreading.  The top two curves assume emission from
the near edge of the jet (equation \ref{eqn:withjetandspreading}) 
with $v=c$ and $v=0$ respectively,
while the bottom curve assumes emission only from the jet center
(equation \ref{eqn:nojetnospread}).
\label{thetavariation}}
\end{figure}

Absent the capability to detect emission from viewing angles near
unity, measuring the GRB jet angle using orphans is all but
impossible.  This is shown in Fig. \ref{thetavariation}, in which we
plot the number of expected afterglow detections per year per square
degree as a function of jet opening angle, for a limiting magnitude of
$R=27$.  As mentioned above, there is essentially no variation in the
predicted number of detections as the jet solid angle $\Omj$ is varied
by orders of magnitude, until nearly isotropic jets 
($\thj\gtrsim 30^\circ$) are reached.
Now, once the GRB afterglow
luminosity function is well measured, as should be possible with
future missions such as Swift or HETE-II, and once GRB jet dynamics
are better understood, then a
prediction can be made for the expected number of afterglows any given
survey can detect.  A significant downward departure from this expected rate
would be hard to reconcile with small jet angles, within the context
of accepted models.  A statistically
significant detection of such a deficit would require a very deep survey
expected to detect hundreds of orphans over a large region of sky.

To establish our conclusion of the difficulty of using optical afterglow
searches to constrain small beaming angles, we have presented two sets of
detection rates in an attempt to bracket the effects of uncertain jet
morphology and evolution.  
We caution that our upper bound is probably not realistic
because it assumes that the jet emits all of its flux as near as possible
to the direction of the observer.  In addition, for both the
upper and lower bounds we have neglected the frequency redshift of photons
received by off axis observers relative to on-axis observers. This should
be taken into account for observers interested in the narrow frequency
range within the optical band.  For a
narrow pencil beam moving along the $\theta=0$ axis the relative redshift
of photons received by an observer at $\theta_{obs}$ is
$\nu(\theta_{obs})/\nu(\theta =0)=(1-\beta\cos\theta_{obs})/(1-\beta)$.
If the afterglow spectrum behaves as $F_{\nu}\propto \nu^{-\alpha/2}$
\citep{jets}, then the redshift effect drives the
exponent in equation 3, which we previously found to be $m=4$,
to $m=(4+\alpha/2)$.  Replacing $m=4$ in our equations with
$m=4+\alpha/2$ slightly diminishes the expected number of detections
by $\lesssim 30\%$.  Were an orphan afterglow
search to be performed, these corrections, as well as a detailed study
of the jet evolution and emission, would be required to calculate a
more accurate estimate of the expected rate.

In our calculations we have assumed particular values for the
parameters describing GRB afterglows. These parameters are the power
law exponents $\alpha$ and $\mu$, the absolute magnitude of the
on-axis afterglow at break, and the break time $\tb$. Uncertainties in
these parameters will introduce uncertainties in the overall
rates. For example, changing $\alpha$ from 2.4 to 1.8 increases the
detection rate by a factor of $\sim 3$.  Similarly, changing the
absolute magnitude at break from $M=-25$ to $M=-29$ (an increase in
luminosity by a factor of 40), increases the overall detection rate by
a factor of 4-5, as expected from equation
\ref{eqn:nojetnospread}. Changing $\tb$ does not affect our
calculations for continuous monitoring, but can
change the detection efficiency depending upon the monitoring time.
In the limit of small efficiency the rate should scale as $\tb$.
Most importantly, the constancy of the calculated detection rate with
respect to the jet opening angle remains valid for $\thj\lesssim
20^\circ$.

One caveat to our estimate for the orphan afterglow detection rate and 
to our constraint on the observed ratio of orphans to GRBs
is the possibility of processes that lead
to weakly beamed optical emission while remaining consistent with the
good agreement between the expected and observed power law decay and
break of the late time afterglow.  For example the GRB central engine
might emit, coincident with an ultra-relativistic jet, a moderately
relativistic spherical shell that leads to isotropic optical
emission. This might occur in central engine models where rotation
leads to small baryon contamination and ultra-relativistic flow along
the rotation axis, but also to ``dirtier'', moderately relativistic flow
away from the rotation axis \citep[e.g. in the collapsar model,][]{woosley}.
Such an isotropic component, if bright enough to be
detected off-axis but faint enough to avoid on-axis detection, would
clearly confound our analysis and allow direct measurement of GRB
collimation.  On the other hand, it seems unlikely that we have
underestimated the off axis optical emission from the jet itself,
because this estimate derives principally from simple relativistic
kinematics. Similarly, the upper limit to the degree of angular
spreading in the jet comes from the maximal assumption that the
lateral expansion velocity is $c$ in the jet comoving rest frame.
Recent numerical hydrodynamics simulations \citep{hydrojet}
of the relativistic jet
dynamics appear to indicate that the afterglow emission
is indeed more tightly beamed in the forward direction than we have
assumed, suggesting that our calculations neglecting spreading are
more realistic than our calculations including lateral spreading.

Finally, we note that radio afterglows are not subject to 
the analysis presented here. 
They do not decay as steeply as their higher frequency
counterparts and remain 
visible until the afterglow emission is effectively isotropic.
\citet{loebradio} have proposed that a full-sky
survey sensitive to 0.1 mJy could legitimately test the jet model of
GRB emission.  It appears that this may be the best way to implement
\citeauthor{rhoads97}' suggestion of using burst-less afterglows to 
measure the GRB jet angle.

\acknowledgments{We thank Kev Abazajian, George Fuller, Bob Kehoe,
and Rick Rothschild for helpful discussions.  
This work was supported in part by the
U.S. Department of Energy under grant DEFG03-90-ER 40546.
JP was supported in part by the NSF under grant PHY98-00980.
ND was also supported by the ARCS Foundation.}

\twocolumn


\begin{thebibliography}{}
\bibitem[Frail et al.(2001)]{frail01} Frail, D. A., et al. 2001,
	submitted to Nature, preprint astro-ph/0102282.
\bibitem[Fynbo et al.(2001)]{fynbo} Fynbo, J. U. et al., to appear in
	\aap, preprint astro-ph/0101425.
\bibitem[Granot et al.(2001)]{hydrojet} Granot, J., et al. 2001,
	astro-ph/0103038. 
\bibitem[Greiner et al.(1999)]{xt2} Greiner, J., Hartmann, D., Voges, W., 
	Boller, T., Schwarz, R. \& Zharikov, S. V. 1999, \aap, 353, 998.
\bibitem[Grindlay(1999)]{xt1} Grindlay, J. E. 1999, \apj, 510, 710.
\bibitem[Kulkarni et al.(2000)]{kulkarni} Kulkarni, S., et al. 2000, 
	astro-ph/0002168.
\bibitem[Kumar(1999)]{kumar} Kumar, P. 1999, \apj, 523, L113.
\bibitem[MacFadyen \& Woosley(1999)]{woosley} MacFadyen, A. I. \& Woosley, 
	S. E. 1999, \apj, 524, 262.
\bibitem[M\'esz\'aros, Rees \& Wijers(1999)]{mrw} M\'esz\'aros, P.,
	Rees, M. J. \& Wijers, R. A. M. J. 1999, New Astr., 4(4), 303.
\bibitem[Paciesas et al.(1999)]{batse} Paciesas, W. S., et al. 1999,
	\apjs, 122, 465.
\bibitem[Perlmutter et al.(1999)]{scp} Perlmutter, S., et al. 
	(The Supernova Cosmology Project) 1999, \apj, 517, 565.
\bibitem[Perna \& Loeb(1998)]{loebradio} Perna, R. \& Loeb, A. 1998,
	\apj, 509, L85.
\bibitem[Piran(1999)]{piran} Piran, T. 1999, Phys. Rep., 314, 575.
\bibitem[Rees(1999)]{rees} Rees, M. J. 1999, \aap S, 138, 491.
\bibitem[Rhoads(1997)]{rhoads97} Rhoads, J. E. 1997, \apj, 487, L1.
\bibitem[Rhoads(1999)]{rhoads99} Rhoads, J. E. 1999, \apj, 525, 737.
\bibitem[Rhoads(2000)]{rhoads00} Rhoads, J. E. 2000, submitted to \apj,
	astro-ph/0008461.
\bibitem[Rhoads(2001)]{rhoads01} Rhoads, J. E. 2001, astro-ph/0103028.
\bibitem[Rybicki \& Lightman(1979)]{radbook} Rybicki, G. B. \&
	Lightman, A. P. 1979, {\it Radiative Processes in
	Astrophysics}, (Wiley: New York).
\bibitem[Sari, Piran \& Halpern(1999)]{jets} Sari, R., Piran, T. \&
	Halpern, J.P. 1999, \apj, 519, L17.
\bibitem[Schmidt et al.(1998)]{hzss} Schmidt, B. P. et al. (High-$z$
	Supernova Search) 1998, \apj, 507, 46. 
\bibitem[Steidel et al.(1999)]{steidel} Steidel, C. C., Adelberger, K. L., 
	Giavalisco, M., Dickinson, M., \& Pettini, M. 1999, \apj, 519, 1.
\bibitem[Woods \& Loeb(1998)]{loebsn} Woods, E. \& Loeb, A. 1998,
	\apj, 508, 760.
\bibitem[Woods \& Loeb(1999)]{woodsandloeb1999} Woods, E. \& Loeb, A. 1999,
	\apj, 523, 187.
\bibitem[Wijers et al. (1998)]{wijers} Wijers, R.A.M.J., Bloom, J.S., 
	Bagla, J.S., \& Natarajan, P. 1998, MNRAS, 294, 13.

\end{thebibliography}
\end{document}